\documentclass[12pt,dvips]{article}
\usepackage{rotating}
\usepackage{axodraw}
\usepackage{epsfig}
\usepackage{color}
\usepackage{cite}
\definecolor{GreenYellow}   {cmyk}{0.15,0,0.69,0}
\definecolor{Yellow}        {cmyk}{0,0,1,0}
\definecolor{Goldenrod}     {cmyk}{0,0.10,0.84,0}
\definecolor{Dandelion}     {cmyk}{0,0.29,0.84,0}
\definecolor{Apricot}       {cmyk}{0,0.32,0.52,0}
\definecolor{Peach}         {cmyk}{0,0.50,0.70,0}
\definecolor{Melon}         {cmyk}{0,0.46,0.50,0}
\definecolor{YellowOrange}  {cmyk}{0,0.42,1,0}
\definecolor{Orange}        {cmyk}{0,0.61,0.87,0}
\definecolor{BurntOrange}   {cmyk}{0,0.51,1,0}
\definecolor{Bittersweet}   {cmyk}{0,0.75,1,0.24}
\definecolor{RedOrange}     {cmyk}{0,0.77,0.87,0}
\definecolor{Mahogany}      {cmyk}{0,0.85,0.87,0.35}
\definecolor{Maroon}        {cmyk}{0,0.87,0.68,0.32}
\definecolor{BrickRed}      {cmyk}{0,0.89,0.94,0.28}
\definecolor{Red}           {cmyk}{0,1,1,0}
\definecolor{OrangeRed}     {cmyk}{0,1,0.50,0}
\definecolor{RubineRed}     {cmyk}{0,1,0.13,0}
\definecolor{WildStrawberry}{cmyk}{0,0.96,0.39,0}
\definecolor{Salmon}        {cmyk}{0,0.53,0.38,0}
\definecolor{CarnationPink} {cmyk}{0,0.63,0,0}
\definecolor{Magenta}       {cmyk}{0,1,0,0}
\definecolor{VioletRed}     {cmyk}{0,0.81,0,0}
\definecolor{Rhodamine}     {cmyk}{0,0.82,0,0}
\definecolor{Mulberry}      {cmyk}{0.34,0.90,0,0.02}
\definecolor{RedViolet}     {cmyk}{0.07,0.90,0,0.34}
\definecolor{Fuchsia}       {cmyk}{0.47,0.91,0,0.08}
\definecolor{Lavender}      {cmyk}{0,0.48,0,0}
\definecolor{Thistle}       {cmyk}{0.12,0.59,0,0}
\definecolor{Orchid}        {cmyk}{0.32,0.64,0,0}
\definecolor{DarkOrchid}    {cmyk}{0.40,0.80,0.20,0}
\definecolor{Purple}        {cmyk}{0.45,0.86,0,0}
\definecolor{Plum}          {cmyk}{0.50,1,0,0}
\definecolor{Violet}        {cmyk}{0.79,0.88,0,0}
\definecolor{RoyalPurple}   {cmyk}{0.75,0.90,0,0}
\definecolor{BlueViolet}    {cmyk}{0.86,0.91,0,0.04}
\definecolor{Periwinkle}    {cmyk}{0.57,0.55,0,0}
\definecolor{CadetBlue}     {cmyk}{0.62,0.57,0.23,0}
\definecolor{CornflowerBlue}{cmyk}{0.65,0.13,0,0}
\definecolor{MidnightBlue}  {cmyk}{0.98,0.13,0,0.43}
\definecolor{NavyBlue}      {cmyk}{0.94,0.54,0,0}
\definecolor{RoyalBlue}     {cmyk}{1,0.50,0,0}
\definecolor{Blue}          {cmyk}{1,1,0,0}
\definecolor{Cerulean}      {cmyk}{0.94,0.11,0,0}
\definecolor{Cyan}          {cmyk}{1,0,0,0}
\definecolor{ProcessBlue}   {cmyk}{0.96,0,0,0}
\definecolor{SkyBlue}       {cmyk}{0.62,0,0.12,0}
\definecolor{Turquoise}     {cmyk}{0.85,0,0.20,0}
\definecolor{TealBlue}      {cmyk}{0.86,0,0.34,0.02}
\definecolor{Aquamarine}    {cmyk}{0.82,0,0.30,0}
\definecolor{BlueGreen}     {cmyk}{0.85,0,0.33,0}
\definecolor{Emerald}       {cmyk}{1,0,0.50,0}
\definecolor{JungleGreen}   {cmyk}{0.99,0,0.52,0}
\definecolor{SeaGreen}      {cmyk}{0.69,0,0.50,0}
\definecolor{Green}         {cmyk}{1,0,1,0}
\definecolor{ForestGreen}   {cmyk}{0.91,0,0.88,0.12}
\definecolor{PineGreen}     {cmyk}{0.92,0,0.59,0.25}
\definecolor{LimeGreen}     {cmyk}{0.50,0,1,0}
\definecolor{YellowGreen}   {cmyk}{0.44,0,0.74,0}
\definecolor{SpringGreen}   {cmyk}{0.26,0,0.76,0}
\definecolor{OliveGreen}    {cmyk}{0.64,0,0.95,0.40}
\definecolor{RawSienna}     {cmyk}{0,0.72,1,0.45}
\definecolor{Sepia}         {cmyk}{0,0.83,1,0.70}
\definecolor{Brown}         {cmyk}{0,0.81,1,0.60}
\definecolor{Tan}           {cmyk}{0.14,0.42,0.56,0}
\definecolor{Gray}          {cmyk}{0,0,0,0.50}
\definecolor{Black}         {cmyk}{0,0,0,1}
\definecolor{White}         {cmyk}{0,0,0,0}

\newcommand{\lsim}{\stackrel{<}{{}_\sim}}
\newcommand{\gsim}{\stackrel{>}{{}_\sim}}

\begin{document}

\def\thefootnote{\fnsymbol{footnote}}

\begin{flushright}
{\tt MAN/HEP/2009/32}\\
{\tt arXiv:0909.1749} \\
September 2009
\end{flushright}

\begin{center}
{\bf {\LARGE Strangephilic Higgs Bosons in the MSSM} }
\end{center}

\medskip

\begin{center}{\large
J.~S.~Lee$^a$, Y.~Peters$^b$, A.~Pilaftsis$^b$ and C.~Schwanenberger$^b$}
\end{center}

\begin{center}
{\em $^a$Physics Division, National Center for Theoretical Sciences, 
Hsinchu, Taiwan}\\[0.2cm]
{\em $^b$School of Physics and Astronomy, University of Manchester,}\\
{\em Manchester M13 9PL, United Kingdom}
\end{center}

\bigskip\bigskip

\centerline{\bf ABSTRACT}
\noindent  
We  suggest  a  new  CPX-derived  scenario  for  the  search  of  {\it
strangephilic} MSSM Higgs bosons at the Tevatron and the LHC, in which
all neutral and charged Higgs bosons decay predominantly into pairs of
strange quarks  and into  a strange and  a charm  quark, respectively.
The proposed  scenario is realized  within a particular region  of the
MSSM parameter  space and requires large values  of $\tan\beta$, where
threshold  radiative   corrections  are  significant   to  render  the
effective   strange-quark  Yukawa  coupling   dominant.   Experimental
searches  for neutral  Higgs  bosons based  on  the identification  of
$b$-quark jets or $\tau$ leptons  may miss a strangephilic Higgs boson
and  its  existence could  be  inferred  indirectly  by searching  for
hadronically decaying charged  Higgs bosons.  Potential strategies and
experimental challenges  to search  for strangephilic Higgs  bosons at
the Tevatron and the LHC are discussed.

\newpage

\section{Introduction}

The search strategy for Higgs bosons at high-energy colliders, such as
LEP, the  Tevatron, and the LHC, depends  crucially on their decay
properties, and especially  on the strength of their  couplings to the
kinematically                     allowed                     decaying
particles~\cite{Barate:2003sz,Schael:2006cr,Carena:2000yx,Aad:2009wy,
Ball:2007zza}.   In the  Standard Model  (SM), the  Higgs boson Yukawa
couplings to fermions are directly proportional to the fermion masses,
i.e.
\begin{equation}
  \label{hfSM}
h_f^{\rm SM} \ = \ \frac{\sqrt{2}\,m_f}{v}\ ,
\end{equation}
where $v\simeq 246$  GeV is the vacuum expectation  value (VEV) of the
SM  Higgs   doublet.   Thus,  the  bottom-quark   Yukawa  coupling  is
suppressed,  for example,  by  the factor  $m_b/m_t$  compared to  the
top-quark one.   However, in extensions of  the SM with  more than one
Higgs  doublet, the  above relation~(\ref{hfSM})  between  the fermion
mass and its Yukawa coupling  gets modified.  In the so-called minimal
supersymmetric  standard   model  (MSSM)  which   involves  two  Higgs
doublets, one has at the tree-level
\begin{equation}
  \label{eq:SUSY_Yukawa}
h_{f=u}^{\rm MSSM} \ = \ \frac{\sqrt{2}\,m_u}{v\,\sin\beta}\,, \ \ \ \ \ \
h_{f=d\,,l}^{\rm MSSM} \ = \ \frac{\sqrt{2}\,m_{d,l}}{v\,\cos\beta}\,,
\end{equation}
where $\tan\beta=\sin\beta/\cos\beta$ denotes the ratio of the VEVs of
the two  Higgs doublets  and $u,d,l$ stand  for the up-  and down-type
quarks  and   charged  leptons,  respectively.   In   this  case,  the
bottom-quark Yukawa  coupling $h_b$ can  be as large as  the top-quark
one $h_t$ for large values  of $\tan\beta \sim m_t/m_b$.  In contrast, the
ratios of the Yukawa couplings  among the down-type quarks and charged
leptons remain the same as in the SM:
\begin{equation}
  \label{eq:hsb}
\frac{h_{d,s}^{\rm MSSM}}{h_{b}^{\rm MSSM}} \ = \
\frac{h_{d,s}^{\rm SM}}{h_{b}^{\rm SM}} \ = \ \frac{m_{d,s}}{m_b}\,, \
\ \ \ \ \ 
\frac{h_{e,\mu}^{\rm MSSM}}{h_{\tau}^{\rm MSSM}} \ = \
\frac{h_{e,\mu}^{\rm SM}}{h_{\tau}^{\rm SM}} \ = \ \frac{m_{e,\mu}}{m_\tau}\ .
\end{equation}
As a  consequence, the  relative branching ratios  of the  neutral and
charged Higgs bosons into light  fermions do not alter, e.g.~$B(H_{1,2,3} \to
s\bar{s})/B(H_{1,2,3}\to b\bar{b}) \ll  1$ and $B(H^+ \to c\bar{s})/B(H^+\to
\tau^+\nu_\tau) \ll 1$ for large values of $\tan\beta$.

In this  paper we propose a  new benchmark scenario for  the search of
neutral  and charged  Higgs bosons,  $H_{1,2,3}$ and  $H^\pm$,  at the
Tevatron and  the LHC.  In this  scenario all the  neutral and charged
Higgs  bosons  decay predominantly  into  strange  quarks  and into  a
strange and a charm quarks,  respectively.  As we will show in Section
2, this scenario occurs within a restricted area of the MSSM parameter
space characterized  by large  values of $\tan\beta$,  where threshold
radiative  corrections are  significant  to give  rise  to a  sizeable
effective strange-quark  Yukawa coupling $h_s$  of order 1\footnote{We
note that our scenario differs from that in~\cite{Carena:1999bh} where
some of the Higgs bosons may favorably decay into light quarks. In our
case,  instead,  {\it all}  charged  and  neutral  Higgs bosons  decay
predominantly to  strange quarks thanks to a  large effective $h_s$.}.
Because  of their  strong affinity  to strange  quarks, we  call these
Higgs scalars {\it strangephilic}.  The particular strangephilic Higgs
scenario, which we  study in detail in Section~3,  is characterized by
the  fact  that  $B(H_{1,2,3} \to  s\bar{s})/B(H_{1,2,3}\to  b\bar{b})
\stackrel{>}{{}_\sim}    1$   and   $B(H^+    \to   c\bar{s})/B(H^+\to
\tau^+\nu_\tau) \stackrel{>}{{}_\sim} 1$ and  by large decay widths of
order 10~GeV for the Higgs bosons $H_{1,2,3}$ and $H^\pm$.  In Section
4,  we discuss  potential strategies  and the  associated experimental
challenges to look for  strangephilic neutral and charged Higgs bosons
at  the  Tevatron and  the  LHC.  Finally,  the  main  results of  our
analysis are summarized in Section~5.

\section{Strangephilic Higgs Bosons}

In the MSSM, the tree-level relations between the masses and couplings
given in  (\ref{eq:SUSY_Yukawa}) get significantly  modified by finite
loop-induced threshold corrections mediated by the exchange of gluinos
and  charginos~\cite{ref:Threshold}.  For  the down-type  quark Yukawa
couplings, such  a modification becomes important for  large values of
$\tan\beta$.   More  explicitly, at  large  $\tan\beta$ the  down-type
quark Yukawa couplings take on the form
\begin{equation}
h_q\ =\ \frac{\sqrt{2}\,m_q}{v\,\cos\beta}\ \frac{1}{1+\Delta_q\,\tan\beta}
\end{equation}
where  $q=d,s,b$  and~\footnote{Throughout this  work,  we follow  the
  conventions and notations of {\tt CPsuperH}~\cite{cpsuperh}.}
\begin{eqnarray}
\Delta_d \!&=&\! \frac{2\alpha_s}{3\pi}\,\mu^*M_3^*\,
I(M_{\tilde{Q}_1}^2,M_{\tilde{D}_1}^2,|M_3|^2)\,, \nonumber \\
\Delta_s \!&=&\! \frac{2\alpha_s}{3\pi}\,\mu^*M_3^*\,
I(M_{\tilde{Q}_2}^2,M_{\tilde{D}_2}^2,|M_3|^2)\,, \nonumber \\
\Delta_b \!&=&\! \frac{2\alpha_s}{3\pi}\,\mu^*M_3^*\,
I(M_{\tilde{Q}_3}^2,M_{\tilde{D}_3}^2,|M_3|^2)
\ + \ \frac{|h_t|^2}{16\pi^2}\ \mu^* A_t^*
I(M_{\tilde{Q}_3}^2,M_{\tilde{U}_3}^2,|\mu|^2)\,,
\end{eqnarray}
with the loop function $I(a,b,c)$ defined as
\begin{equation}
I(a,b,c)\ =\
\frac{ab\ln(a/b) + bc\ln(b/c) + ac\ln(c/a)}
{(a-b)(b-c)(a-c)}\ .
\end{equation}
Considering $\alpha_s  \sim 0.1$ and  $|h_t| \sim 1$,  the gluino-loop
contribution  to  $\Delta_b$  is  estimated  to  be  larger  than  the
chargino-loop  one by  the factor  $\sim \pi  |\mu|^2/|A_t\,M_3|$ when
$|\mu|^2\,,|M_3|^2\,\gg\, M^2_{\tilde{Q}_3,\tilde{U}_3,\tilde{D}_3}$
\cite{Borzumati:2004rd}.
In   this   case,   assuming   the  universal   sfermion   masses   or
$M_{\tilde{Q}_3}=M_{\tilde{Q}_2}$,   $M_{\tilde{U}_3}=M_{\tilde{U}_2}$,
and   $M_{\tilde{D}_3}=M_{\tilde{D}_2}$,  one  has   $\Delta_s  \simeq
\Delta_b$ leading  to the same relation~(\ref{eq:hsb}) as  in the case
without the inclusion of threshold corrections, i.e.  $B(H_{1,2,3} \to
s\bar{s})/B(H_{1,2,3}\to   b\bar{b})    \ll   1$   and    $B(H^+   \to
c\bar{s})/B(H^+\to \tau^+\nu_\tau) \ll 1$.
However,  for  particular   choices  of  the  theoretical  parameters,
$\Delta_s$ can differ dramatically from $\Delta_b$.  In particular, if
$|1+\Delta_s \tan\beta| \to 0$  and $|1+\Delta_b \tan\beta| \sim {\cal
O}(1)$~\cite{DP},  this would  result in  a scenario  with  $|h_s| \gg
|h_b|$,  where  all  neutral   Higgs  bosons  $H_{1,2,3}$  will  decay
predominantly  into  strange  quarks  and the  charged  Higgs  scalars
$H^\pm$ into strange and charm quarks.

In  our  phenomenological  analysis,  we will  utilize  the  effective
Lagrangian  approach to describe  the interaction  of the  neutral and
charged Higgs bosons to quarks in the presence of $\tan\beta$-enhanced
threshold corrections.  The effective Lagrangian is given by
\begin{eqnarray}
{\cal L}_{Hqq^\prime} =
-\sum_{i} g_q H_i \bar{q}
(g^S_{_{H_i\bar{q}q}}+ig^P_{_{H_i\bar{q}q}}\gamma_5)q
-\left[g_{q^\prime q} H^{+} \bar{q^\prime}
(g^S_{_{H^+\bar{q^\prime}q}}+ig^P_{_{H^+\bar{q^\prime}q}}\gamma_5)\, q
\ \ +{\rm h.c.}\right]\,,
\end{eqnarray}
where $q=d,s,b$ and $q^\prime=u,c,t$.
At the tree level, 
\begin{equation}
g_q=\frac{g\,m_q}{2\,M_W}\,, \ \ \
g^S_{_{H_i\bar{q}q}}=O_{\phi_1\,i}/\cos\beta\,, \ \ 
g^P_{_{H_i\bar{q}q}}=-\tan\beta\,O_{a\,i}\,
\end{equation}
for the  neutral Higgs boson  $H_{1,2,3}$ with the $3\times  3$ mixing
matrix      $O$      such     that      $(\phi_1,\phi_2,a)^T=O_{\alpha
i}\,(H_1,H_2,H_3)^T$  with $M_{H_1} \leq  M_{H_2} \leq  M_{H_3}$.  For
the charged Higgs boson,
\begin{equation}
g_{q^\prime q}\ =\ -\frac{g\,m_{q^\prime}}{\sqrt{2}\,M_W}\,, \ \ \
g^S_{_{H^+\bar{{q^\prime}}q}}\ =\
\frac{1}{2}\left[\frac{1}{\tan\beta}+\frac{m_q}{m_{q^\prime}}\,
  \tan\beta\right]\,, \ \ \ 
g^P_{_{H^+\bar{{q^\prime}}q}}\ =\
\frac{i}{2}\left[\frac{1}{\tan\beta}-\frac{m_q}{m_{q^\prime}}\,
  \tan\beta\right]\,. 
\end{equation}
In   the   presence    of   threshold   corrections,   the   couplings
$g^S_{_{H_i\bar{q}q}}$,                         $g^P_{_{H_i\bar{q}q}}$,
$g^S_{_{H^+\bar{q^\prime}q}}$     and    $g^P_{_{H^+\bar{q^\prime}q}}$
read~\cite{Pilaftsis:2002fe}:
\begin{eqnarray}
g^S_{H_i\bar{q}q} & =& {\rm Re}\, \bigg(\,
\frac{1}{1\, +\, \Delta_q\,\tan\beta}\,\bigg)\,
\frac{O_{\phi_1 i}}{\cos\beta}
\ +\ {\rm Re}\, \bigg(\, \frac{\Delta_q}{1\, +\,
\Delta_q\, \tan\beta}\,\bigg)\
\frac{O_{\phi_2 i}}{\cos\beta}
\nonumber\\ &&
+\: {\rm Im}\, \bigg[\,
\frac{ \Delta_q\, (\tan^2\beta\, +\, 1)}{1\, +\,
\Delta_q\, \tan\beta}\,\bigg]\
O_{ai}\, , \nonumber\\[0.35cm]
g^P_{H_i\bar{q}q} & =& -\, {\rm Re}\, \bigg(\,
\frac{ \tan\beta\, -\, \Delta_q}{1\, +\, \Delta_q \tan\beta}\,\bigg)\, O_{ai}
\ +\ {\rm Im}\, \bigg(\, \frac{\Delta_q\,\tan\beta}{1\, +\,
\Delta_q\, \tan\beta}\,\bigg)\
\frac{O_{\phi_1 i}}{\cos\beta}\nonumber\\
&&-\: {\rm Im}\, \bigg(\,
\frac{\Delta_q}{1\, +\, \Delta_q\, \tan\beta}\,\bigg)\
\frac{O_{\phi_2 i}}{\cos\beta}\ , 
\label{eq:nhqq} \\[0.35cm]
g^S_{_{H^+\bar{q^\prime}q}} &=&
\frac{1}{2}\left[\frac{1}{\tan\beta}+\frac{m_q}{m_{q^\prime}}\
\frac{\tan\beta\,-\,\Delta_q^*}{1\,+\,\Delta_q^*\,\tan\beta}\right]\,,
\nonumber \\ 
g^P_{_{H^+\bar{q^\prime}q}} &=&
\frac{i}{2}\left[\frac{1}{\tan\beta}-\frac{m_q}{m_{q^\prime}}\
\frac{\tan\beta\,-\,\Delta_q^*}{1\,+\,\Delta_q^*\,\tan\beta}\right]\,.
\label{eq:chqq}
\end{eqnarray}
For  completeness, the  charged Higgs boson  couplings to  leptons are
given by
\begin{equation}
g_{\nu_l l}=-\frac{g\,m_{l}}{\sqrt{2}\,M_W}\,, \ \ \
g^S_{_{H^+\bar{\nu}_ll}}=\tan\beta/2\,, \ \ \
g^P_{_{H^+\bar{\nu}_ll}}=-i\,\tan\beta/2\,.
\end{equation}
In the limit of $|1+\Delta_s^*  \tan\beta| \to 0$, it is not difficult
to       show       that      $|g^{S,P}_{H_{1,2,3}\bar{s}s}|       \gg
|g^{S,P}_{H_{1,2,3}\bar{b}b}|$    and   $|g^{S,P}_{H^+\bar{c}s}|   \gg
|g^{S,P}_{H^+\bar{\nu}_\tau\tau}|$. For Higgs  bosons lighter than the
top quark,  this possibility leads  to a strangephilic  scenario where
the  neutral and charged  Higgs bosons  will decay  predominantly into
$s\bar{s}$  and $c\bar{s}$,  respectively, instead  of  $b\bar{b}$ and
$\tau^+\nu_\tau$.
In our  numerical analysis, we have  included the effects  of the mass
splitting  in  third  generation   squarks  and  all  other  threshold
corrections that are not enhanced by $\tan\beta$~\cite{cpsuperh}.

Finally,   it  is  important   to  remark   that  the   proposed  {\em
  strangephilic}    scenario    will    generically    induce    large
flavour-changing-neutral-current  (FCNC)  effects  mediated  by  Higgs
bosons at large $\tan\beta$, because of the inherent hierarchy between
the first  two and third  generations~\cite{DP}.  One possible  way to
avoid these large FCNC effects would  be to go beyond the framework of
minimal  Supergravity (mSUGRA) and  allow for  sizeable flavour-mixing
effects  in  the squark  sector,  such  that  a kind  of  cancellation
mechanism   becomes  operative~\cite{DP}.   Evidently,   FCNC  effects
strongly depend  on the strength  of the {\em  off-diagonal} effective
Yukawa couplings and the flavour structure of the model in general. We
will  not address  this highly  model-dependent issue  in  the present
work. Our interest here is to analyze the implications of a large {\em
  diagonal} strange-quark effective  Yukawa coupling, independently of
the {\em off-diagonal} ones,  for Higgs-boson searches at the Tevatron
and the~LHC.

\section{Strangephilic Higgs Benchmark Scenarios}

As strangephilic viable models, we consider the following two
benchmark scenarios derived from CPX~\cite{Carena:2000ks}:
\begin{eqnarray}
{\bf L}~:~ && \hspace{-0.6cm}
M_{\tilde{Q}_3} = M_{\tilde{U}_3} = M_{\tilde{D}_3} = 0.5\,{\rm TeV}\,;\,
M_{\tilde{L}_3}= M_{\tilde{E}_3} = 0.7\,{\rm TeV}\,,
\nonumber \\ && \hspace{-0.6cm}
M_{H^\pm}=130\,{\rm GeV}\,, \ \
|\mu|=2\,{\rm TeV}\,, \ \
|A_{t,b,\tau}|=1\,{\rm TeV} \,, \ \
|M_3|=1 ~~{\rm TeV}\,,
\nonumber \\ && \hspace{-0.6cm}
\Phi_{A_t}=\Phi_{A_b}=\Phi_{A_\tau}=90^\circ\,, \ \
\Phi_3=180^\circ\,,
\nonumber \\ && \hspace{-0.6cm}
\rho_{\tilde{Q}}= \rho_{\tilde{U}}= \rho_{\tilde{L}}= \rho_{\tilde{E}}=1\,, 
\nonumber \\ && \hspace{-0.6cm}
1\leq\tan\beta\leq 120\,, \ \ 2\leq \rho_{\tilde{D}} \leq 6\,,
\label{eq:CPX_L} 
\\[0.3cm]
{\bf S}~:~ && \hspace{-0.6cm}
M_{\tilde{Q}_3} = M_{\tilde{D}_3} = 2\,{\rm TeV}\,;\,
M_{\tilde{U}_3} = M_{\tilde{L}_3}= M_{\tilde{E}_3} = 1\,{\rm TeV}\,,
\nonumber \\ && \hspace{-0.6cm}
M_{H^\pm}=120\,{\rm GeV}\,, \ \
|\mu|=2\,{\rm TeV}\,, \ \
|A_{t,b,\tau}|=1\,{\rm TeV} \,, \ \
|M_3|=1 ~~{\rm TeV}\,,
\nonumber \\ && \hspace{-0.6cm}
\Phi_{A_t}=\Phi_{A_b}=\Phi_{A_\tau}=90^\circ\,, \ \
\Phi_3=180^\circ\,,
\nonumber \\ && \hspace{-0.6cm}
\nonumber \rho_{\tilde{U}}= \rho_{\tilde{L}}= \rho_{\tilde{E}}=1\,,
\\ && \hspace{-0.6cm}
1\leq\tan\beta\leq 60\,, \ \ 0.1\leq \rho_{\tilde{Q},\tilde{D}} \leq 0.6\,.
\label{eq:CPX_S}
\end{eqnarray}
In the above, we  have introduced the $\rho_{\tilde{X}}$ parameters to
account  for  a  mass  hierarchy  between  the  first  two  and  third
generations:
\begin{equation}
M_{\tilde{X}_{1,2}}=\rho_{\tilde{X}}\, M_{\tilde{X}_3}
\end{equation}
with     $\tilde{X}=\tilde{Q},\tilde{U},\tilde{D},\tilde{L},\tilde{E}$.
In  the {\bf  L} scenario,  the Higgs  bosons exhibit  a strangephilic
behaviour for {\bf large} values of $\rho_{\tilde{D}}$ greater than 1,
whereas in the {\bf S}  scenario the strangephilic nature of the Higgs
bosons     is     realized    for     {\bf     small}    values     of
$\rho_{\tilde{Q},\tilde{D}}$  less  than  1.  Finally,  the  remaining
parameters are fixed as
\begin{eqnarray}
&&|M_1|=250~{\rm GeV}\,, \ \ \ \Phi_1=0^\circ\,; \ \ \
|M_2|=500~{\rm GeV}\,, \ \ \ \Phi_2=0^\circ\,; \ \ \
\nonumber \\
&&|A_{e,\mu,u,c,d,s}|=|A_{t,b,\tau}|\,, \ \ \
|\Phi_{A_{e,\mu,u,c,d,s}}|=0^\circ\,, \ \ \ \Phi_\mu=0\,.
\label{eq:others}
\end{eqnarray}

\begin{figure}[H]
\hspace{ 0.0cm}
\vspace{-0.5cm}
\centerline{\epsfig{figure=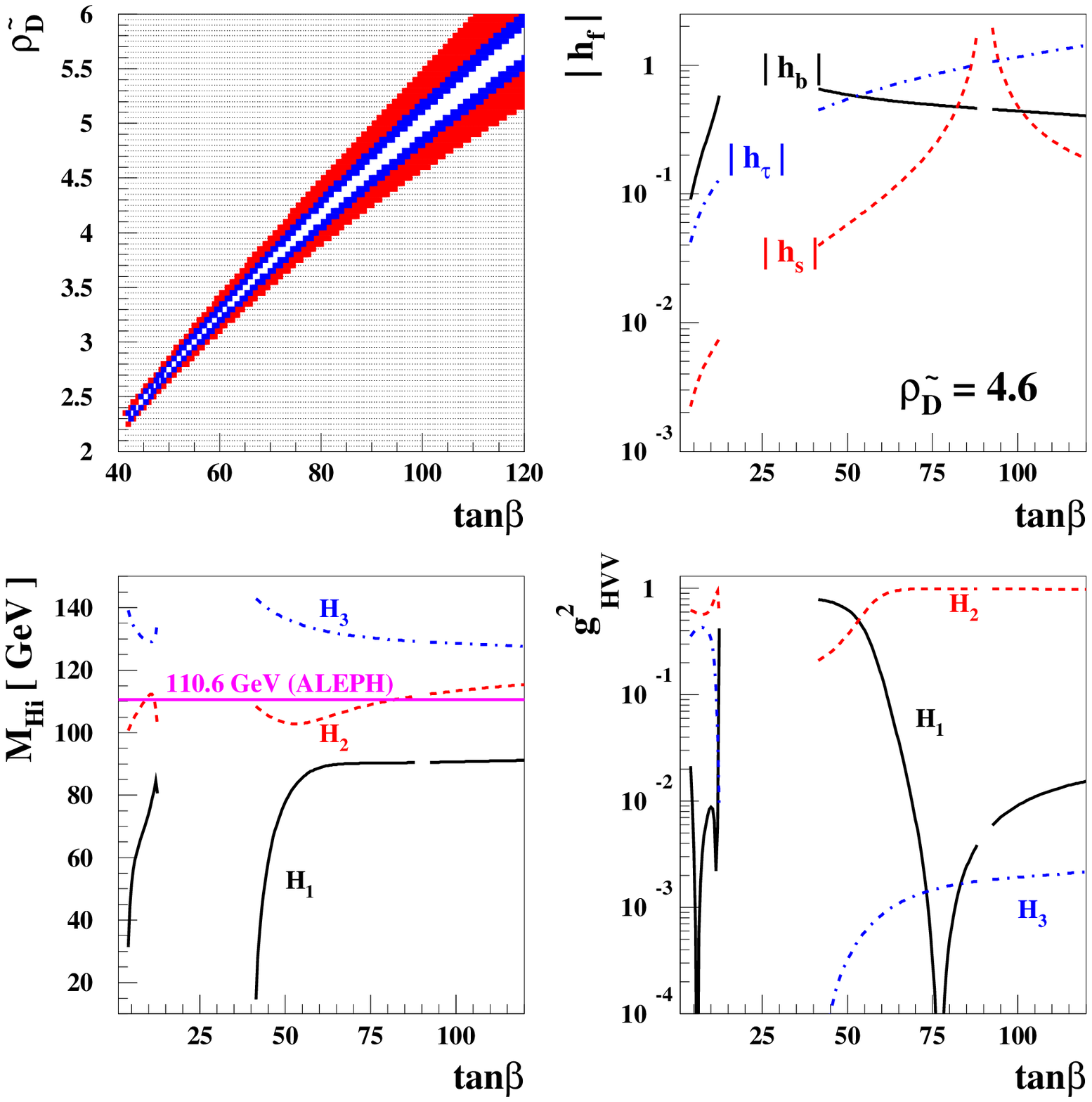,height=14.0cm,width=14.0cm}}
\caption{\it  Numerical  estimates in  the  {\bf  L} scenario  defined
  in~(\ref{eq:CPX_L}).   The upper-left frame  shows the  region (blue
  and    red),    in    which    $|h_s|    \geq    |h_b|$    in    the
  $\rho_{\tilde{D}}$-$\tan\beta$  plane.  The  narrow  unshaded region
  sandwiched between  the shaded regions is  not allowed theoretically
  resulting in  a non-perturbative value  $|h_s| > 2$, while  the red-
  and blue-shaded  regions correspond  to $|h_s| \leq  1$ and  $1 \leq
  |h_s| \leq 2$, respectively.  In the upper-right frame, the absolute
  values of  the bottom- and  strange-quark and the  tau-lepton Yukawa
  couplings are shown as functions of $\tan\beta$ in the solid, dashed
  and       dash-dotted       lines,      respectively,       assuming
  $\rho_{\tilde{D}}=4.6$. The region  between $\tan\beta=12$ and 41 is
  not allowed theoretically giving rise to tachyonic bottom squarks or
  to a Higgs boson mass  matrix with a complex or negative eigenvalue.
  The lower frames  display the masses and their  couplings to the $W$
  or $Z$ bosons ($V=W,Z$) as functions of $\tan\beta$.}
\label{fig:rhodtbL}
\end{figure}

In the  upper-left frame of  Fig.~\ref{fig:rhodtbL}, the blue  and red
regions show  where the strange-quark  Yukawa coupling is  larger than
the  bottom-quark  Yukawa one,  which  is  obtained by  simultaneously
varying               the               hierarchy               factor
$\rho_{\tilde{D}}=M_{\tilde{D}_{1,2}}/M_{\tilde{D}_3}$              and
$\tan\beta$.  The unshaded region  is not theoretically allowed, as it
leads  to  a  non-perturbative  value  for  the  strange-quark  Yukawa
coupling where $|h_s|>2$
\footnote{This constraint is close to the limit for which 
the Yukawa coupling becomes non-perturbative at the
renormalization-group scale of 1-2  TeV, i.e.~ $|h_s|^2/{4\pi}\,>\,1$.}.
The  larger $\rho_{\tilde{D}}$  is,  the
larger  value   for  $\tan\beta$  is  needed  to   satisfy  the  limit
$|1+\Delta_s^*  \tan\beta| \to  0$  and so  realize the  strangephilic
condition: $|h_s|  > |h_b|$.   In the upper-right  frame, we  show the
absolute values  of the bottom-  and strange-quark and  the tau-lepton
Yukawa     couplings    as     functions    of     $\tan\beta$,    for
$\rho_{\tilde{D}}=4.6$. In  the low $\tan\beta$ region  we always have
$|h_s| \ll |h_b|$.  As  $\tan\beta$ increases, the bottom-quark Yukawa
coupling grows  rapidly and, in the region  between $\tan\beta=12$ and
41,  results in  tachyonic bottom  squarks  or in  a Higgs-boson  mass
matrix with  a complex or  negative eigenvalue.  For larger  values of
$\tan\beta$, in  the region  $80 \lsim \tan\beta  \lsim 100$,  one can
have $|h_s| > |h_b|$,  where $|1+\Delta_s^* \tan\beta| \simeq 0$.  The
unshaded region  in the upper-left  frame corresponds to  the interval
$88 \lsim \tan\beta \lsim 92$,  within which $|h_s|>2$ lies beyond the
realm of  validity of perturbation theory.   We note that  in the high
$\tan\beta$ region, we  can always have $|h_s| >  |h_b|$ for any value
of   $\tan\beta$   by   appropriately   tuning  the   free   parameter
$\rho_{\tilde{D}}$ as shown in the upper-left frame.

\begin{figure}[!htb]
\hspace{ 0.0cm}
\vspace{-0.5cm}
\centerline{\epsfig{figure=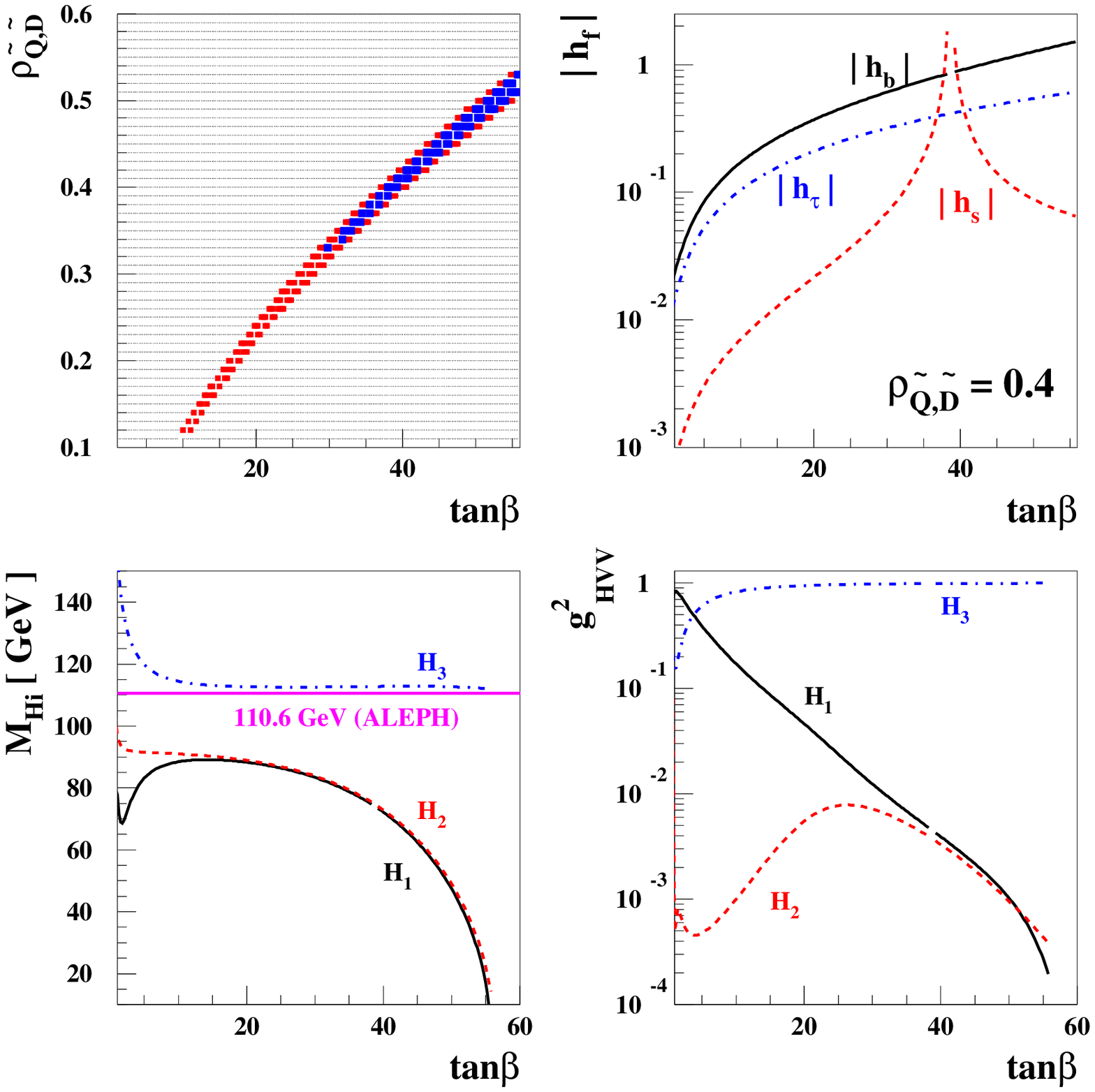,height=14.0cm,width=14.0cm}}
\caption{\it The  same as in Fig.~\ref{fig:rhodtbL}, but  for the {\bf
    S}  scenario defined  in~(\ref{eq:CPX_S}).  The  upper  left frame
  shows the region, for  which $|h_s| \geq |h_\tau|$.  The upper-right
  frame displays the  dependence of the absolute values  of the Yukawa
  couplings       $|h_{s,b,\tau}|$      on       $\tan\beta$,      for
  $\rho_{\tilde{Q},\tilde{D}}=0.4$.}
\label{fig:rhodtbS}
\end{figure}
In the lower-left frame of Fig.~\ref{fig:rhodtbL}, 
we show the three neutral Higgs boson masses as 
functions of $\tan\beta$ together with the 110.6 GeV (95 \% C.L.)
experimental bound on
the flavor-independent hadronically decaying SM Higgs boson derived
from LEP~\cite{Heister:2002cg}
\footnote{Conservatively, we require that the mass of a {\em
    strangephilic} Higgs boson be larger than 110.6 GeV, when its
  coupling to the $Z$ boson is larger than about 0.1 of the SM
  $HZZ$-coupling.
 A more  precise treatment  of the LEP  and TEVATRON limits  might be
  obtained      by      using      the      public      code      {\tt
    HiggsBounds}~\cite{Bechtle:2008jh}.}.    For   large   values   of
$\tan\beta$,  the lightest  Higgs  boson lies  below the  experimental
bound  but  escapes  the  LEP   searches  due  to  its  small  couping
$g_{H_1VV}^2 \lsim  0.02$, if  $\tan\beta \gsim 65$,  as shown  in the
lower-right frame.  In this kinematic region, the coupling of $H_2$ to
the gauge bosons is large, $g_{H_2VV}^2 \gsim 0.97$.  However, for $80
\stackrel{<}{{}_\sim} \tan\beta  \stackrel{<}{{}_\sim} 100$, the $H_2$
boson  is  predominantly   strangephilic  with  a  mass  $M_{H_2}\gsim
110$~GeV, so  it remains undetected thanks  to the flavour-independent
LEP limit mentioned above.  For $\tan\beta \stackrel{>}{{}_\sim} 100$,
the $H_2$ boson is heavier than $\sim 114$~GeV, thereby satisfying the
absolute LEP bound on the  SM Higgs boson.  Thus, for $\tan\beta \gsim
80$ and $\rho_{\tilde{D}} > 4$, any strangephilic region is compatible
with the LEP limits in the {\bf L}~benchmark scenario (\ref{eq:CPX_L}).
In this respect, we should comment that the existence of uncertainties
due  to possible variations  of the  top-quark mass,  the higher-order
quantum     corrections      and     the     known     field-theoretic
differences~\cite{Higgs:Uncertainty}  between the Feynman-diagrammatic
and RG-improved approaches may  result in significant uncertainties in
the  predictions for  the MSSM  Higgs-boson mass  spectrum,  which are
typically  bigger  than  $\sim  3$~GeV.  Therefore,  given  all  these
different sources  of uncertainties,  the constraints derived  here on
the Higgs-boson  masses and their couplings  to the Z  boson using the
public code {\tt CPsuperH}, which implements the RG-improved approach,
should be regarded as fair and conservative.

One can  obtain a  similar LEP2-compatible strangephilic  scenario for
smaller  values  of $\tan\beta$  in  the  {\bf  S} benchmark  scenario
(\ref{eq:CPX_S}), for $\rho_{\tilde{Q},\tilde{D}} < 1$.
Specifically,  the upper-left  frame  of Fig.~\ref{fig:rhodtbS}  shows
that $|h_s|$ can be larger  than $|h_\tau|$ for $30 \lsim \tan\beta \lsim
55$, when the free  parameter $\rho_{\tilde{Q},\tilde{D}}$ is tuned to
a relative narrow region around  a particular value between $\sim 0.3$
and  $\sim   0.5$.   For  instance,   in  the  upper-right   frame  of
Fig.~\ref{fig:rhodtbS},  we see  that $|h_s|$  is dominant  within the
narrow region: $37  \lsim \tan\beta \lsim 40$, for  the specific value
of $\rho_{\tilde{Q},\tilde{D}}  = 0.4$.   We note that  for $\tan\beta
\gsim 30$ and $\rho_{\tilde{Q},\tilde{D}}  \gsim 0.3$, the two lighter
Higgs   bosons   cannot  be   detected   by   LEP  searches,   because
$g^2_{H_{1,2}VV} \lsim 0.01$, as can  be seen from the lower frames of
Fig.~\ref{fig:rhodtbS}.   Moreover, the mass  of the  heaviest neutral
Higgs state $H_3$ is always beyond the current LEP limit.

Even though  the proposed benchmark  scenarios are not  generic within
the MSSM, the  degree of parameter tuning required  for realizing {\em
strangephilic} Higgs  bosons is not excessive.  As we will  see in the
next section, there is a  significant range of $\tan\beta$ values, for
which strangephilic Higgs bosons can occur within the {\bf L} and {\bf
S} benchmark scenarios.

We conclude this section by noticing that both the {\bf L} and {\bf S}
scenarios   require    large   CP   phases   for    the   $A$   terms,
i.e.~$\Phi_{A_t}=\Phi_{A_b}=\Phi_{A_\tau}=90^\circ$,   in   order   to
satisfy   the  flavour-independent   LEP  bound~\cite{Heister:2002cg}.
These   large  CP   phases   together  with   the   large  $|\mu|=   2
|A_{t,b,\tau}|$   are   the  characteristic   features   of  the   CPX
scenario~\cite{Carena:2000ks}.   There are  severe constraints  on the
size of the CP-odd phases  which arise from the non-observation of the
Thallium,    neutron,   and    Mercury    electric   dipole    moments
(EDMs). Specifically,  we note that the  enhanced strange-quark Yukawa
coupling may  induce large neutron EDM  in the Parton  Quark Model and
large Higgs-mediated EDMs.   Nevertheless, these constraints could, in
principle,  be   evaded  in  CPX-like  scenarios,   by  arranging  for
cancellations among  the different one- and  two-loop contributions to
EDMs~\cite{Pilaftsis:2002fe,Ellis:2008zy}.

\section{Potential Search Strategies and Experimental\\ Challenges}

\begin{figure}[!htb]
\hspace{ 0.0cm}
\vspace{-0.5cm}
\centerline{\epsfig{figure=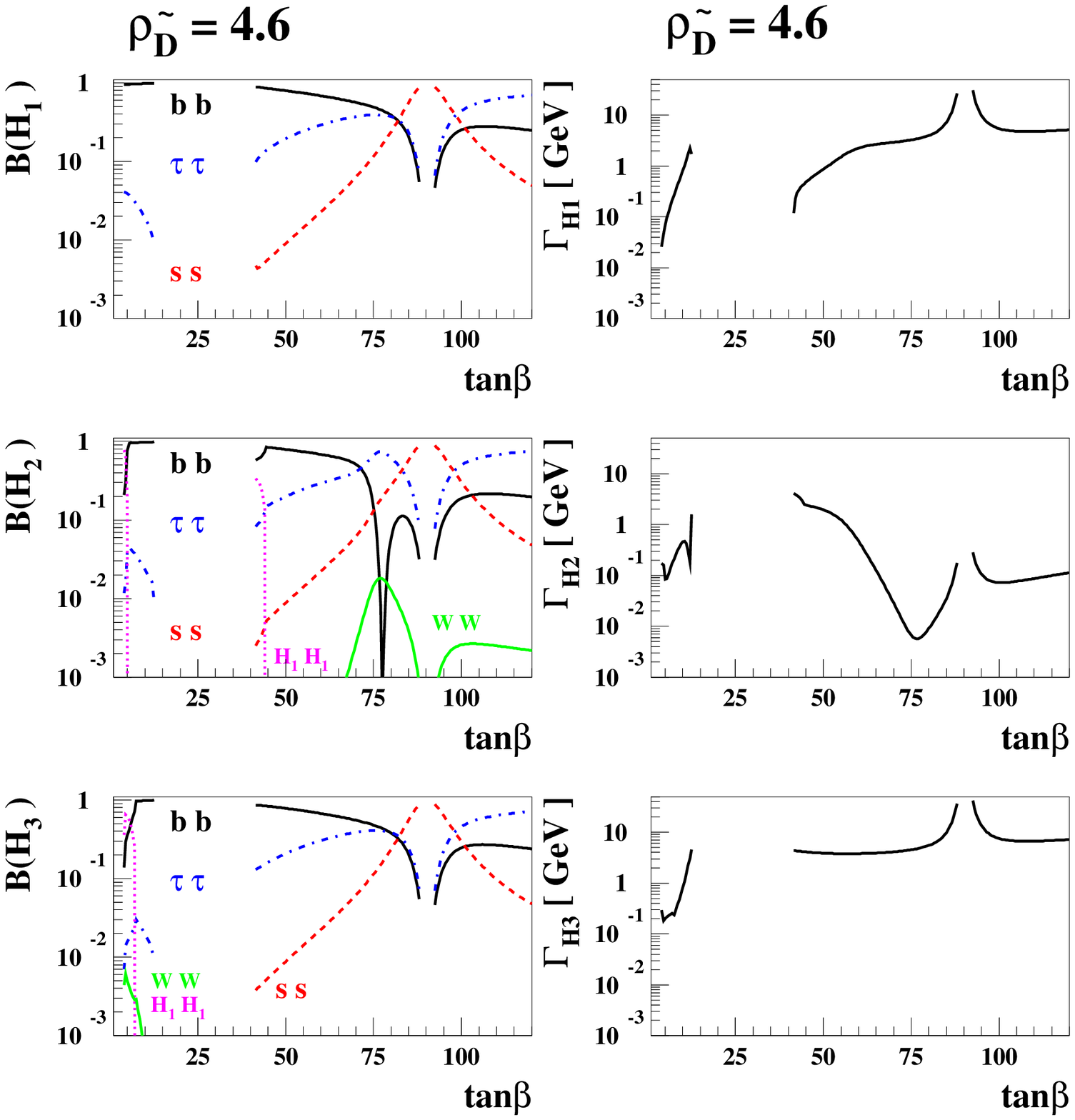,height=14.0cm,width=14.0cm}}
\caption{\it Numerical estimates of branching ratios (right) and total
  decay widths (left) of the three neutral Higgs bosons $H_{1,2,3}$ as
  functions of $\tan\beta$, for $\rho_{\tilde{D}}=4.6$, in the {\bf L}
  scenario given in~(\ref{eq:CPX_L}).  }
\label{fig:nhL}
\end{figure}
\begin{figure}[!htb]
\hspace{ 0.0cm}
\vspace{-0.5cm}
\centerline{\epsfig{figure=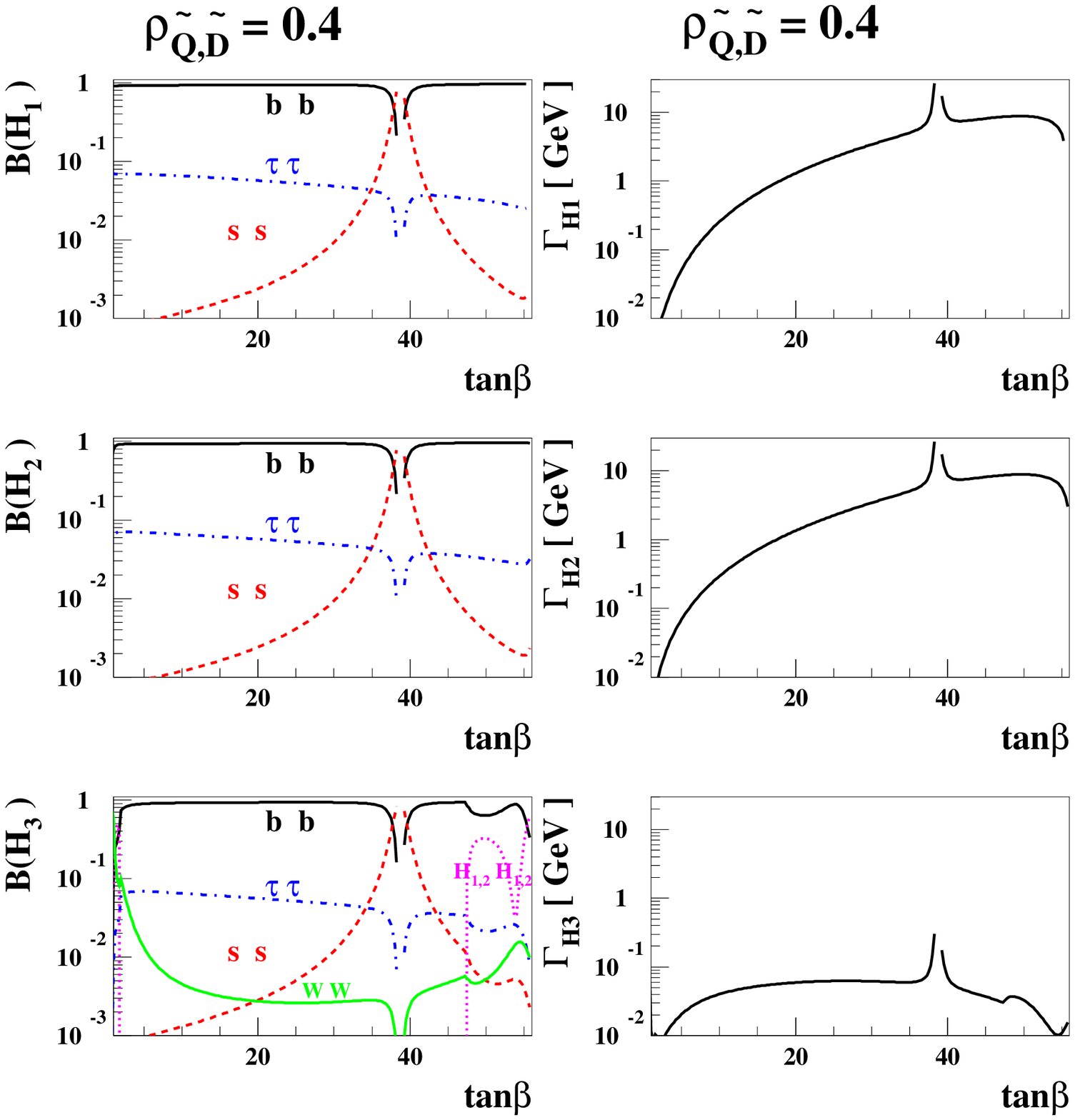,height=14.0cm,width=14.0cm}}
\caption{\it Numerical estimates of branching ratios (right) and total
  decay widths (left) of the three neutral Higgs bosons $H_{1,2,3}$ as
  functions  of $\tan\beta$, for  $\rho_{\tilde{Q},\tilde{D}}=0.4$, in
  the {\bf S} scenario given in~(\ref{eq:CPX_S}).  }
\label{fig:nhS}
\end{figure}
 
In this  section, we analyze the generic  phenomenological features of
the strangephilic Higgs  bosons in the {\bf L}  and {\bf S} scenarios.
We   also  discuss  potential   search  strategies   and  experimental
challenges  to  look  for  strangephilic  neutral  and  charged  Higgs
scalars.

Let us first consider  the neutral Higgs sector.  Figure~\ref{fig:nhL}
shows our numerical results for  the three neutral Higgs bosons in the
{\bf L} scenario.   We observe that it is always  possible to choose a
LEP2-compatible $\rho_{\tilde{D}}$ to realize a strangephilic scenario
for    any   large    value   of    $\tan\beta   >    80$.    Choosing
$\rho_{\tilde{D}}=4.6$,   we   display   in   Fig.~\ref{fig:nhL}   the
dependence of the  branching ratios (left frames) and  the total decay
widths (right frames)  on $\tan\beta$.  We note the  large total decay
widths of  the order of a few  GeV or larger where  $B(H_{1,2,3} \to s
\bar{s})  > B(H_{1,2,3}  \to  b \bar{b})$  and/or  $B(H_{1,2,3} \to  s
\bar{s}) > B(H_{1,2,3} \to \tau^+ \tau^-)$.
The  width  of the  $H_2$  boson  is not  as  enhanced  as those  of
$H_{1,3}$ and remains  below $0.3$ GeV, since the  $H_2$ boson is most
likely $\phi_2$, which does not couple to the down-type quarks and the
charged leptons at the tree level.

Figure  \ref{fig:nhS}  shows our  numerical  results  in  the {\bf  S}
benchmark scenario,  for $\rho_{\tilde{Q},\tilde{D}} =  0.4$.  In this
case, one can get a strangephilic scenario around $\tan\beta \sim 38$;
see the  upper-right frame  of Fig.~\ref{fig:rhodtbS}.  As  before, we
observe that the decay widths of the $H_{1,2}$ bosons are large ${\cal
  O}(10)$~GeV,  whereas the $H_3$  width $\Gamma_{H_3}$  remains below
$0.3$~GeV,  since the  $H_3$ boson  is  most likely  $\phi_2$ in  this
scenario.

The strangephilic region, $80 \lsim \tan\beta \lsim 100$ ($35 \lsim 
\tan\beta \lsim 42$)  for the {\bf L} ({\bf S})  scenario with 
$\rho_{\tilde{D}}=4.6$ ($\rho_{\tilde{Q},\tilde{D}} = 0.4$), is determined 
by the condition that $B(H_{1,2,3} \to s \bar{s})$ is bigger than 
$B(H_{1,2,3} \to b \bar{b})$ or bigger than $B(H_{1,2,3} \to \tau^+ 
\tau^-)$.  
At the Tevatron and the LHC,
strangephilic  neutral Higgs bosons
can be substantially produced through  the fusion of strange 
quarks: $s\bar{s}  \to H_{1,2,3}$. The Higgs
bosons produced in the $s$-channel would decay into $s$ quark and,
without any efficient tagging technique for the identification 
of $s$-quark jets, the observation of strangephilic neutral Higgs bosons 
is experimentally challenging, if not impossible, unless
the neutral Higgs bosons have sizeable branching ratios to photons.
In particular, conventional searches 
at the Tevatron or the LHC that rely on tagging of $b$-quark jets or the 
identification of $\tau$ leptons will miss a strangephilic neutral Higgs 
boson.  

Alternatively,  one may consider  searching for  strangephilic neutral
Higgs  bosons  in  associate  production  channels, such  as  $gg  \to
t\bar{t} H_{1,2,3}$,  $qq \to qq  H_{1,2,3}$ and $qq^{(\prime)}  \to V
H_{1,2,3}$,  with $V=Z,W^\pm$.  However,  conventional searches  at the
Tevatron and the LHC for  such processes assume that the neutral Higgs
bosons  decay predominantly into  $b\bar{b}$, $\tau^+\tau^-$,  
or $W^+ W^-$.
Again,  without  an  efficient $s$-quark  jet
tagging  identification,  the  light-quark  multijet  background  will
render the sensitivity of  the conventional search strategies to those
channels  rather   problematic.
One viable option might be to perform a flavour-independent 
analysis of the central exclusive diffractive Higgs production, $p p \to p 
+ H_{1,2,3} + p$, which could help to determine the mass and the broad 
width of the strangephilic neutral Higgs bosons, provided $H_{1,2,3}$ are 
produced at sufficiently high rates.


%
\begin{figure}[!htb]
\hspace{ 0.0cm}
\vspace{-0.5cm}
\centerline{\epsfig{figure=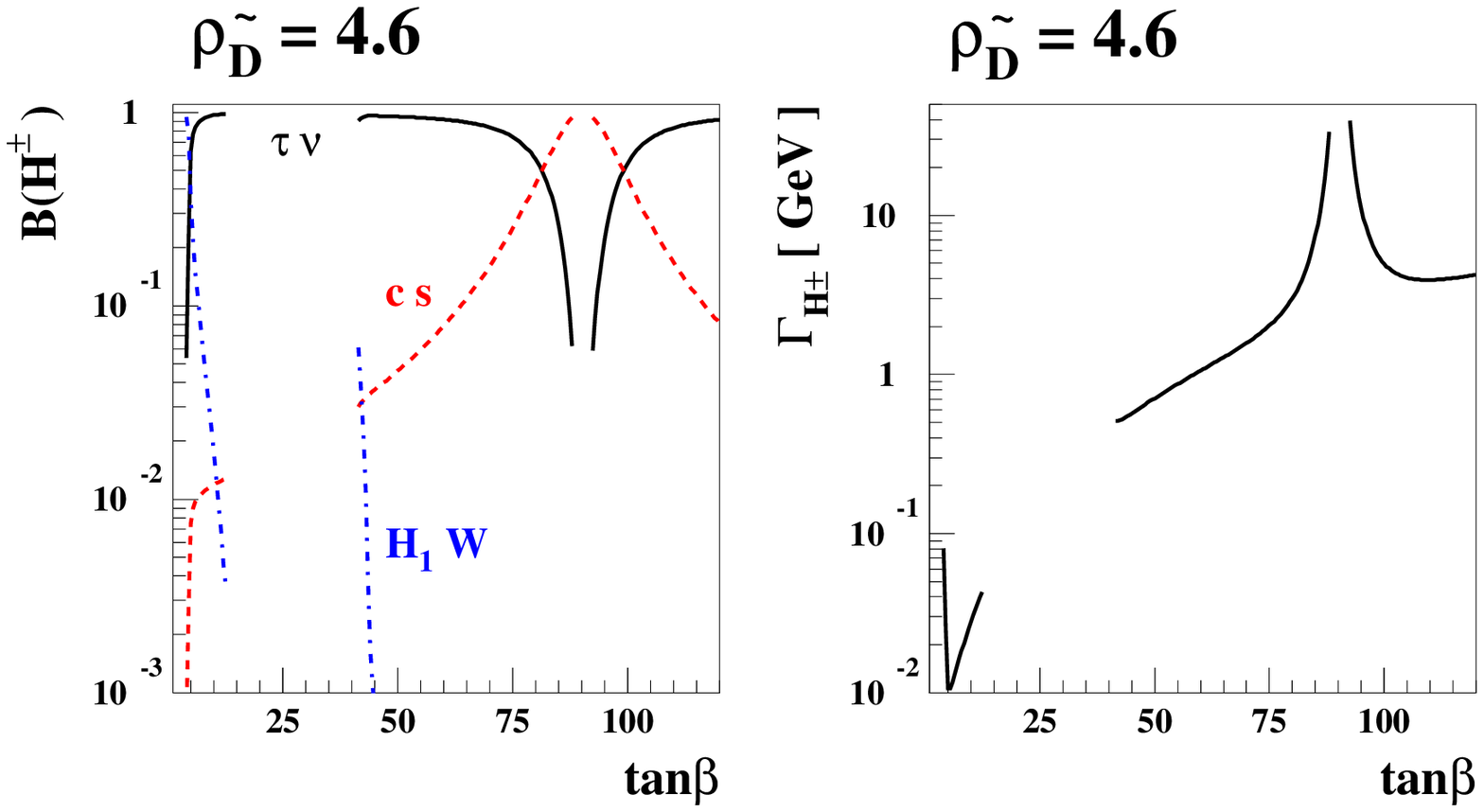,height=7.0cm,width=14.0cm}}
\caption{\it
The same as in Fig.~\ref{fig:nhL} but for the charged Higgs boson.  }
\label{fig:chL}
\end{figure}
\begin{figure}[!htb]
\hspace{ 0.0cm}
\vspace{-0.5cm}
\centerline{\epsfig{figure=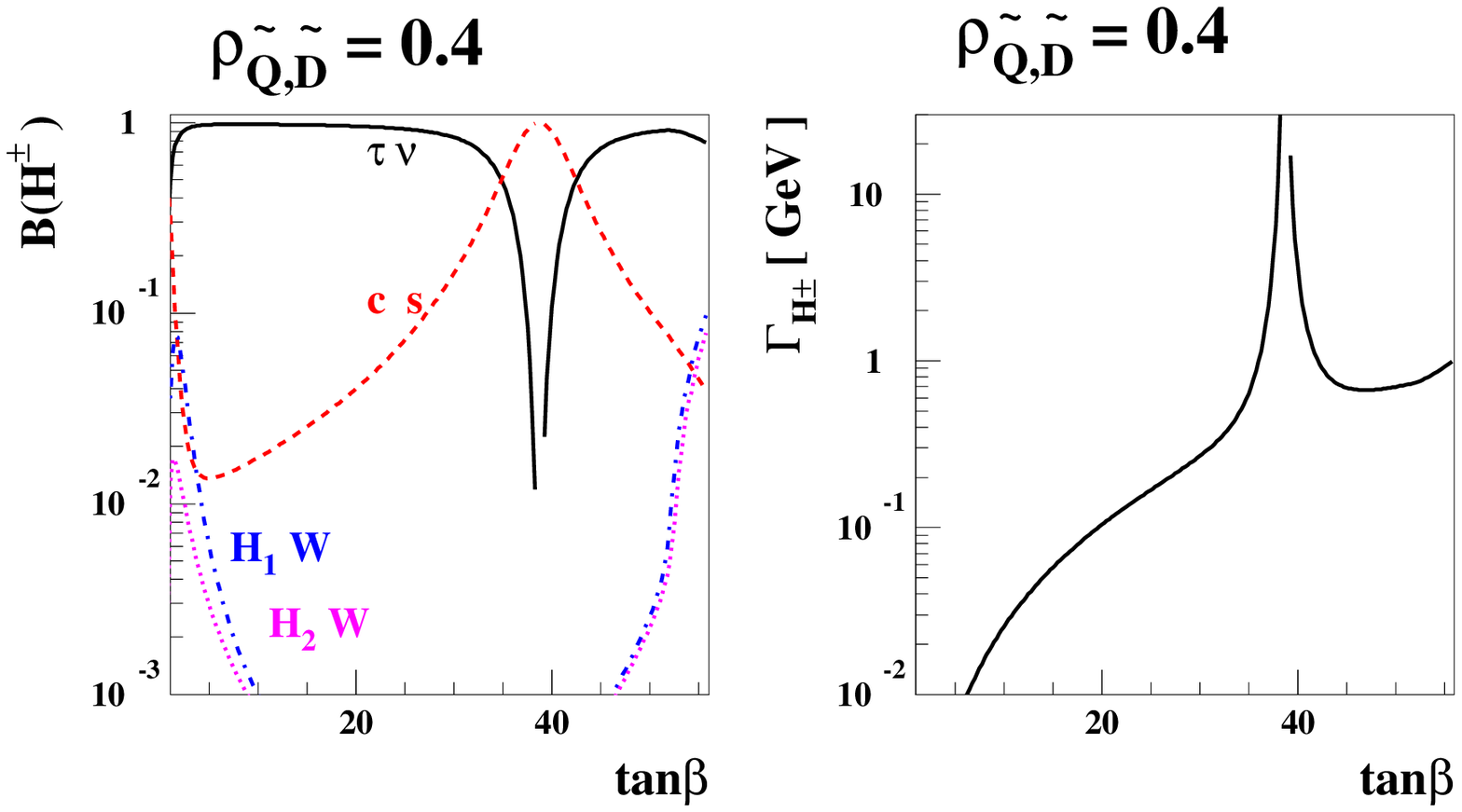,height=7.0cm,width=14.0cm}}
\caption{\it
The same as in Fig.~\ref{fig:nhS} but for the charged Higgs boson.  }
\label{fig:chS}
\end{figure}

\begin{figure}[!htb]
\hspace{ 0.0cm}
\vspace{-0.5cm}
\centerline{\epsfig{figure=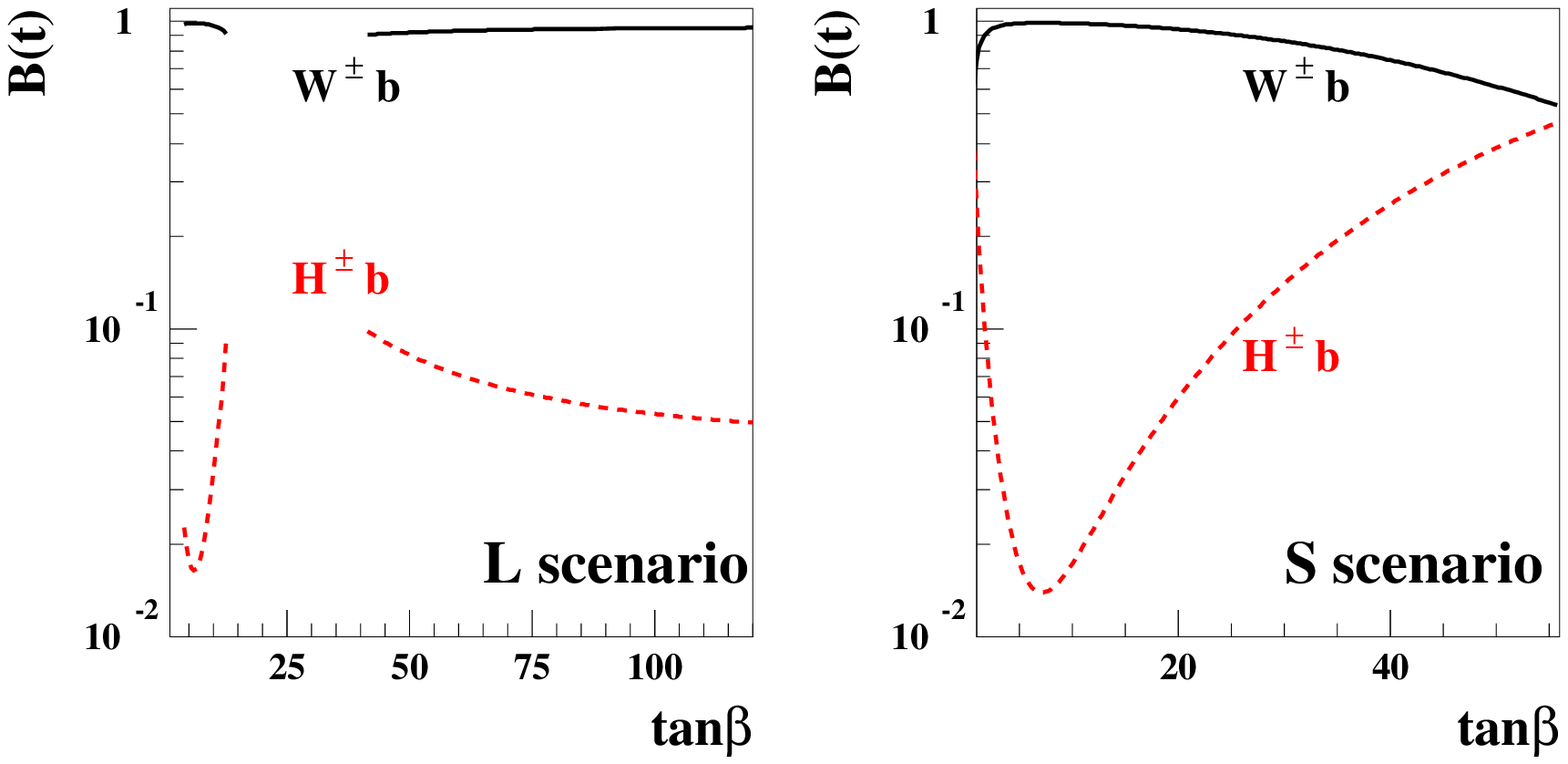,height=7.5cm,width=14.0cm}}
\caption{\it The top-quark branching ratios  in the {\bf L} (left) and
  {\bf   S}  (right)   scenarios,   defined  in~(\ref{eq:CPX_L})   and
  (\ref{eq:CPX_S}), respectively.  }
\label{fig:topbr}
\end{figure}

Let  us   now  turn  our   attention  to  the  charged   Higgs  boson.
Figures~\ref{fig:chL} and \ref{fig:chS} show numerical results for the
charged  Higgs   boson  in  the   {\bf  L}  and  {\bf   S}  scenarios,
respectively.
Like the neutral Higgs bosons, strangephilic  charged  Higgs bosons
can be copiously produced through  the fusion of strange and charm
quarks, $c\bar{s} \to H^+$. But, their observation 
would be challenging, requiring the detection of $s$ and $c$ quarks.
Unlike  the   neutral  Higgs  bosons, however,
light  strangephilic charged  Higgs  bosons can  be
searched  for in  the  $t\bar{t}$ production  channel
at  the  Tevatron and  the  LHC, where the  top
quarks decay  subsequently into $H^\pm$ and  $b$-quarks, provided $B(t
\to H^+  b)$ is substantial~\cite{Abazov:2009ae}.  
Including threshold
corrections, the partial decay width $t \to H^+ b$ is given by
\footnote{The calculation  of the top-quark branching  ratios has been
implemented  in  the  most  recent  version of  {\tt  CPsuperH2.0}  by
including         the         ${\cal         O}(\alpha_s)$         QCD
corrections~\cite{Chetyrkin:1999br},   as   well   as  the   threshold
corrections.}
\begin{eqnarray}
  \label{BtHb}
\Gamma(t\rightarrow H^+ b)&=&\frac{g_{tb}^2m_t}{16\pi}
\left(|g^S_{_{H^+\bar{t}b}}|^2+|g^P_{_{H^+\bar{t}b}}|^2\right)
\left(1-\frac{M_{H^\pm}^2}{m_t^2}\right)^2\,.
\end{eqnarray}
As shown in Fig.~\ref{fig:topbr}, the branching ratio~$B(t \to H^+ b)$
is smaller than 10\% in the {\bf L} scenario, whilst  it can be much
larger  in the  {\bf  S} scenario  depending  on the  actual value  of
$\tan\beta$.

We  note that  conventional  searches for  charged  Higgs bosons  that
analyze only the tauonic  decay channel $H^+\to \tau^+\nu$ loose their
sensitivity in  particular regions  of $\tan\beta$. This  happens, for
example, if $\tan\beta$ is in  the region between $\sim$ 80 and $\sim$
100 in  the {\bf L}  benchmark scenario (see  Fig.~\ref{fig:chL}, left
panel) or  between $\sim$ 35  and $\sim$ 42  in the {\bf  S} benchmark
scenario  (see Fig.~\ref{fig:chS},  left panel).   In both  cases, the
charged Higgs boson decays predominantly  into $c$ and $s$ quarks.  In
order  to cover  the full  range  of $\tan\beta$  without missing  the
possible existence of a  strangephilic charged Higgs boson, both decay
channels,  $H^+\to  c\bar{s}$  and  $H^+\to  \tau^+\nu$,  need  to  be
investigated.

From an experimental point of view, searches for hadronically decaying
charged Higgs bosons, such as strangephilic charged Higgs bosons, have
two major advantages  over the  search for tauonically  decaying charged
Higgs  bosons. 

First, the  detection is  already possible through  the ratios  of the
$t\bar{t}$ production cross  sections in dileptonic, semi-leptonic and
all-hadronic final states.  Compared to the number of $t\bar{t}$ pairs
in  semi-leptonic (dileptonic) final  states, a  strangephilic charged
Higgs  boson would  lead to  a larger  number of  $t\bar{t}$  pairs in
all-hadronic (semi-leptonic)  final states  than predicted by  the SM.
Measurements  of such  cross-section  ratios will  be  available in  a
relative early stage of  the LHC, since large systematic uncertainties
cancel      in       the      ratio.       Recently,       the      D0
collaboration~\cite{Abazov:2009ae}  was  first  to  use  cross-section
ratios to  obtain information on  charged Higgs bosons.   In contrast,
tauonic  charged Higgs  decays lead  to a  disappearance in  all those
channels.  Therefore,  they will  be harder to  detect them  without a
reliable  understanding  of   $\tau$-lepton  identification  which  is
experimentally challenging.

Second,  since a  strangephilic charged  Higgs boson  decays  into two
jets, its full invariant mass  can be reconstructed with a much better
resolution than  in tauonic decays  of the charged Higgs  boson, where
only  the missing transverse  momentum of  the $\tau$-neutrino  can be
used  for  the  mass  reconstruction.   This will  lead  to  a  larger
sensitivity in  the search for strangephilic charged  Higgs bosons.  A
first search for charged Higgs  bosons decaying into charm and strange
quarks with full reconstruction of  the charged Higgs mass was carried
out recently by the CDF collaboration~\cite{CDFHplus}. This recent CDF
analysis can exclude only a very narrow region of the {\bf S} scenario
around $\tan\beta \simeq  38$, provided the mass of  the charged Higgs
boson is small enough.

\bigskip\bigskip
\section{Conclusions}

We have analyzed a new  benchmark scenario that can be realized within
the  MSSM, for  which the  strength of  the $s$-quark  Yukawa coupling
relative  to the $b$-quark  one can  be dramatically  enhanced through
gluino-mediated one-loop corrections.  In particular, for large values
of $\tan\beta$ and for a certain choice of the soft SUSY-breaking mass
parameters, one can obtain a {\em strangephilic} scenario in which the
neutral and  charged Higgs bosons  have dominant or  substantial decay
modes into  a pair of  strange quarks and  into a charm and  a strange
quark, respectively.   Even though not  fully included in  the present
analysis, we note that as well as third generation quarks and squarks,
strange squarks  may also contribute significantly  to the Higgs-boson
mass spectrum through a  large effective strange-quark Yukawa coupling
at the loop level.

At  the  Tevatron collider  and  the  LHC,  strangephilic neutral  and
charged Higgs bosons can be substantially produced through $s$-channel
fusion processes  initiated by strange and charm  quarks.  In general,
their  experimental  observation  would  be  challenging  without  any
efficient tagging technique for  the identification of $s$-quark jets.
As  was  discussed  in   Section~4,  for  the  neutral  Higgs  bosons,
alternative production  channels may be considered,  such as exclusive
diffractive Higgs  production, which could be  helpful in experimental
searches.  On the other hand, it proves easier to detect strangephilic
charged Higgs  bosons $H^\pm$, if produced at  sufficiently high rate,
e.g.~in top-quark decays.  In particular, investigating both the decay
channels  $H^+  \to \tau^+\nu$  and  $H^+  \to  c \bar{s}$  proves  an
important search strategy for  detecting a strangephilic charged Higgs
boson~\cite{Collaboration:2009zh}.    A  positive   signal   in  these
channels  would point  towards an  entire strangephilic  Higgs sector,
which could  be realized even  within a constrained  two-Higgs doublet
model, such as the~MSSM.

In our on-going quest of the thus-far elusive Higgs boson, we strongly
encourage the experimental Collaborations  at the Tevatron and the LHC
to perform a detailed analysis and systematic search for strangephilic
Higgs bosons.

\vspace{-0.2cm}
\section*{Acknowledgements}
\vspace{-0.3cm}
\noindent
We thank  Un-ki Yang for  a discussion concerning  future experimental
prospects  for $s$-quark jet  identification.  The  work of  A.P.  was
supported in part  by the STFC research grant:  PP/D000157/1. The work
of C.S.  and Y.P.  was supported by the Royal Society.

\end{document}